\documentclass[superscriptaddress,12pt,onecolumn,nofootinbib]{revtex4-2} 

\usepackage{amssymb}

\usepackage{mathtools}

\usepackage{tabularray}
\UseTblrLibrary{booktabs}

\usepackage{amsmath}  
\usepackage{amsfonts} 
\usepackage{graphicx} 

\usepackage{xcolor}
\usepackage{appendix}

\usepackage{mathrsfs}

\usepackage{placeins}

\newcounter{researchquestion}

\usepackage[caption=false]{subfig} 

\usepackage{pbox}

\makeatletter
\newcommand{\vast}{\bBigg@{3}}
\newcommand{\Vast}{\bBigg@{3.5}}
\makeatother

\usepackage{hyperref}
\begin{document}


\title{A Shape-Preserving Shell Theorem}


\author{Penelope Messinger}
\thanks{These authors contributed equally to this work.}
\affiliation{Scripps College, Claremont, CA 91711, USA.}

\author{Stephany M. Rodriguez}
\thanks{These authors contributed equally to this work.}
\affiliation{Scripps College, Claremont, CA 91711, USA.}

\author{Trung V. Phan}
\email{{tphan@natsci.claremont.edu}}
\affiliation{Department of Natural Sciences, Scripps and Pitzer Colleges, \\ Claremont Colleges Consortium, Claremont, CA 91711, USA}

\begin{abstract}
We identify all radial interaction potentials that generalize Newton's exterior \textit{shell theorem}, for which a uniformly distributed spherical shell is indistinguishable from a point-source up to \textit{strength renormalization} and \textit{radial rescaling}. We call these \textit{shape-preserving shell interactions}, extending Gurzadyan's cosmological theorem which considers strength renormalization alone. We further expand the theory from \textit{monopolar} to \textit{multipolar} sources and establish the corresponding classification in arbitrary spatial dimensions. We find potential families beyond the \textit{spherical property} subclass and show that the classifications for \textit{multipolar shape-preserving interactions} form a \textit{nested hierarchy}, with each lower-rank class being a subset of the next higher-rank class.
\end{abstract}

\date{\today}

\maketitle 

\section{Introduction}

Newton's \textit{shell theorem} states that, outside a uniform spherical shell distribution of mass, the Newtonian gravitational field is identical to that of a central point mass \cite{newton1833philosophiae,sneddon1949camb,kilmister1974newton}. Gurzadyan's cosmological theorem \cite{gurzadyan1985cosmological,reed2022note,Carimalo2023,phan2025derivation} extends this result to the broader \textit{spherical property} \cite{chapman1983gravity,barnes1984gazette}, identifying all \textit{radial interaction} potentials for which the exterior field is \textit{proportional} to that of a central point mass, with a \textit{strength renormalization} factor that depends only on the shell size. These findings also apply to solid spheres and, more generally, to central interactions whose strength is proportional to an \textit{additive} source quantity, such as electric charge or other conserved charges. Although formulated for \textit{static} fields, analogous sphere-to-point reductions also arise in \textit{dynamical radiative} problems, with \textit{frequency-dependent strength renormalization} \cite{cochran2022introduction}. In our previous work, we have generalized it for \textit{higher dimensions} and \textit{non-trivial topologies} \cite{tse2026generalizing}, using a connection between spherical means and the Euler-Poisson-Darboux (EPD) identity \cite{john2004plane}. A more detailed analysis for the case in flat spaces can be found here \cite{carimalo2026note}.

All the studies mentioned above focus on the \textit{monopole} case, in which the spherical shell is composed of rotationally featureless constituents. This raises a natural question: does an analogous \textit{sphere-to-point reduction} extend beyond the monopole to higher-rank \textit{multipoles}\footnote{A monopole corresponds to a multipole of rank $\ell=0$.}, whose constituents may possess nontrivial angular structure or orientation? Such multipolar structures arise naturally throughout electromagnetism, nuclear physics, gravitation, and radiation theory. Several special cases of \textit{multipolar shell theorem} are known explicitly in three-dimensional space: a uniformly magnetized sphere produces the same exterior field as a central point dipole with the same total dipole moment \cite{edwards2017interactions}, and, more generally, the reduction has been shown to hold for spherically symmetric distributions of aligned multipoles of arbitrary rank for the \textit{inverse-distance} interaction potential \cite{ji2021potential}. However, to the best of our knowledge, a \textit{multipolar spherical property} has not yet been systematically classified across arbitrary dimensions. Another question is whether the \textit{spherical property} can be further generalized by requiring the interaction to retain just the same \textit{shape}, rather than the same radial profile. In other words, the interaction force need only be identical up to not only a \textit{strength renormalization} but also a \textit{radial rescaling}. We call this the \textit{shape-preserving property}.

\begin{figure*}[thbp!]
    \centering
\includegraphics[width=0.9\linewidth]{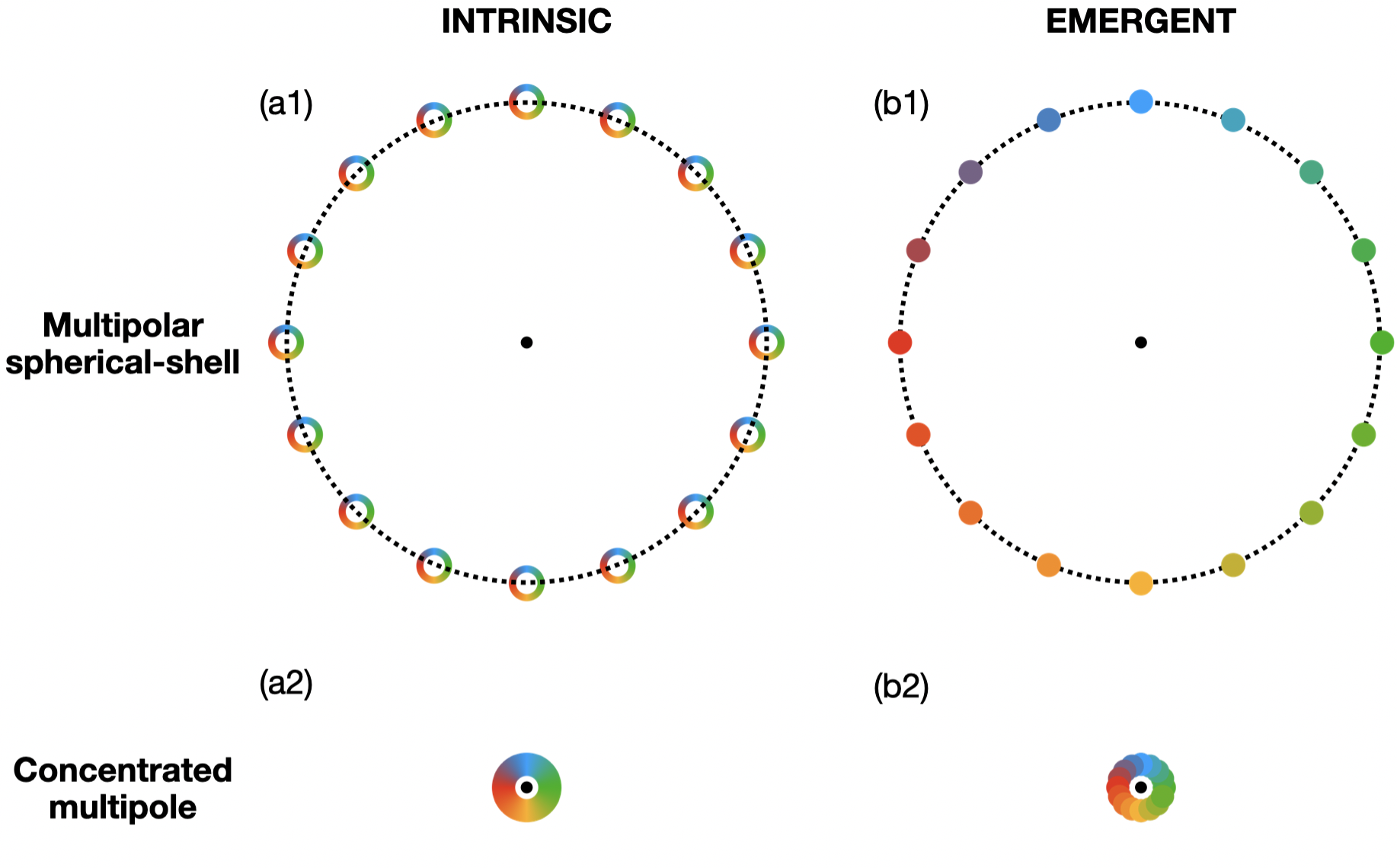}
    \caption{\textbf{The difference between intrinsic multipolarity and emergent multipolarity.} Here, we demonstrate the distinction in two-dimensional space ($n=2$) by representing constituent sources on a circle $S^1$. We only show a finite number of representative point-sources from the continuous shell. \textbf{(a1)} \textit{Intrinsic multipolarity}, in which every constituent is itself a multipole. \textbf{(b1)} \textit{Emergent multipolarity}, in which each constituent is a monopole and the multipolarity arises from their oriented distribution. \textbf{(a2,b2)} The corresponding concentrated versions of the two cases, which become similar.}
\label{fig01}
\end{figure*}

Here, we classify all radial interaction potentials satisfying the \textit{shape-preserving property} for \textit{monopolar} and \textit{multipolar} sources. Section \ref{sec:monopole} considers monopoles, while Section \ref{sec:multipole} extends the classification to multipoles of arbitrary angular rank. We focus on \textit{intrinsic} multipolarity, in which the constituents themselves are multipolar (see Fig. \ref{fig01}a), rather than \textit{emergent} multipolarity arising from an angular distribution of monopoles (see Fig. \ref{fig01}b). For simplicity, we restrict our study to $n$-dimensional flat Euclidean spaces i.e. $\mathbb{R}^n$, so that the multipoles can be aligned by spatial translation.

\section{Monopolar Shape-Preserving Shell Interactions \label{sec:monopole}}

We first formulate the \textit{shape-preserving condition} for a uniform shell of monopoles and use the Euler-Poisson-Darboux (EPD) identity to constrain the radial interaction. We then classify the admissible potentials, identifying \textit{regular} and \textit{singular} families beyond the spherical property and determining their \textit{strength renormalization} and \textit{radial rescaling} factors.

\subsection{The Euler-Poisson-Darboux Identity for Monopolarity}

\begin{figure*}[thbp!]
    \centering
\includegraphics[width=\linewidth]{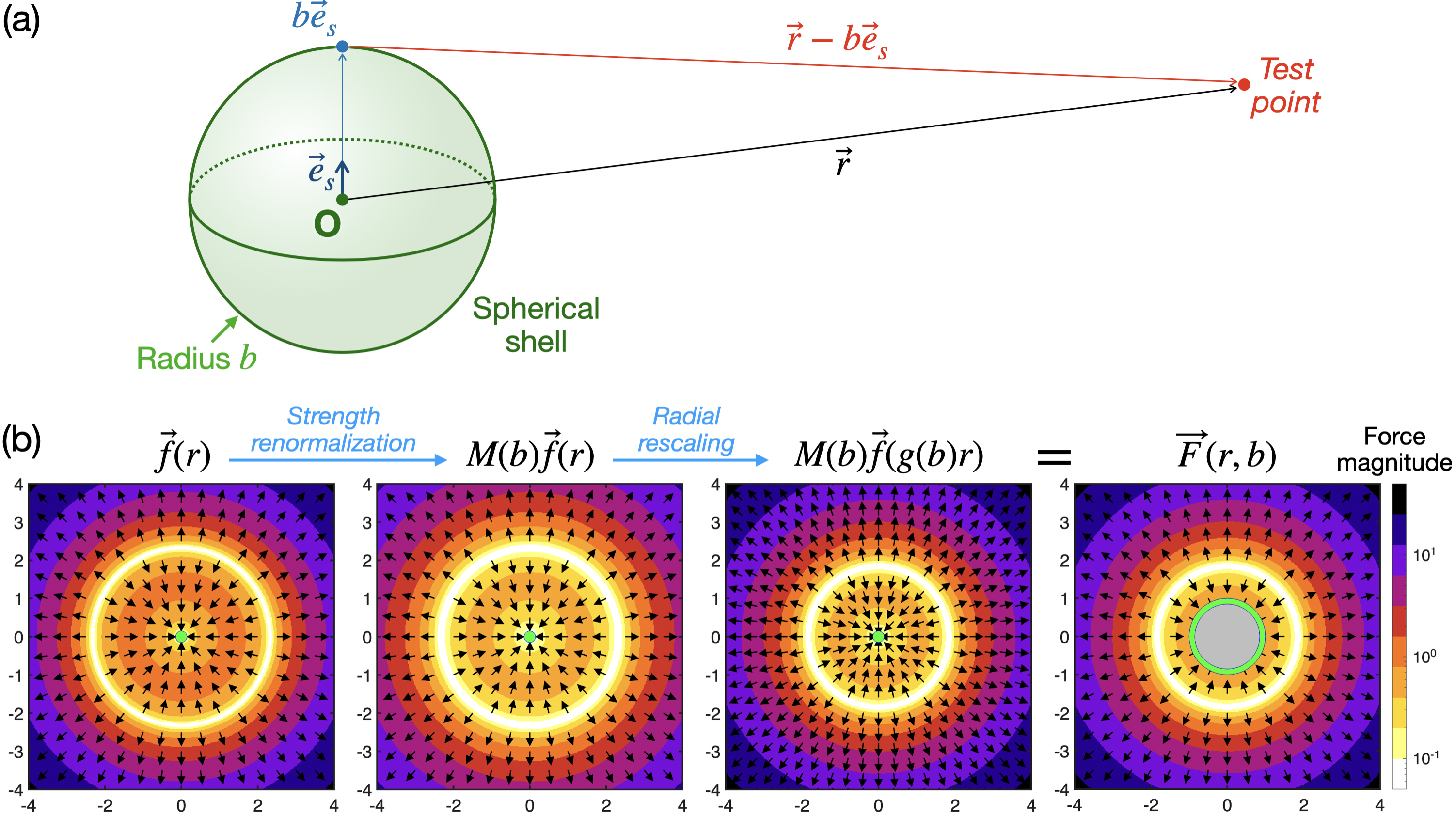}
    \caption{\textbf{Monopolar shape-preservation.} We show the set-up for the spherical shell potential integration Eq. \eqref{eq:full_potential} in \textbf{(a)}, and the mapping of the monopolar point-source force field to that of the uniform spherical shell through \textit{strength renormalization} and \textit{radial rescaling} in \textbf{(b)}. The green dots indicate the point-sources and the hollow green circle represents the spherical shell source, while the \textit{fire}-colormap represents the force magnitude. Arrows indicate the force direction, which is purely radial.}
\label{fig02}
\end{figure*}

For a rigorous treatment, we need to formulate this mathematically. Choose an origin $O\in\mathbb{R}^n$, and consider a point located at $\vec{r}$; in polar coordinates $\vec{r}=(r,\Omega)$, where $\Omega$ specifies the angular direction. Consider a \textit{radial interaction} potential $\phi(\Delta)$ created by a point-source of unit strength at distance $\Delta$, with the source strength assumed to be additive and the potential scaling linearly with the source strength. Outside a uniform spherical shell of radius $b$ consisting of monopoles with total monopolar strength \textit{normalized} to unity, which is centered at the origin $O$, the potential created at position $\vec{r}$ is given by the spherical mean over all shell elements (see Fig. \ref{fig02}a):
\begin{equation}
    \Phi(\vec{r},b) = \frac{ \oint_{S^{n-1}(b)} d\sigma(b \vec{e}_s) \ \phi \left(  | \vec{r}-b\vec{e}_s  | \right)}{\text{Vol}\left[ S^{n-1}(b)\right]} \ ,
\label{eq:full_potential}
\end{equation}
where $S^{n-1}(b)$ is the $(n-1)$-dimensional spherical surface of radius $b$, $\mathrm{Vol}[S^{n-1}(b)]$ is its total hypersurface volume, $\vec{e}_s$ is a unit direction vector on the sphere, $d\sigma(b \vec{e}_s)$ is the surface-area element, and $|\circ |$ denotes the Euclidean norm of a vector quantity $\circ$ (so that $r \equiv|\vec r|>b$ and $|\vec{e}_s|=1$). Another equivalent way to express $\Phi(\vec{r},b)$ in terms of a general function $\phi(\vec{r})$ is through the Pizzetti expansion \cite{sabelfeld1997spherical}:
\begin{equation}
    \Phi(\vec{r},b) = \sum^\infty_{j=0} \frac{\Gamma(n/2)}{4^j j! \Gamma(n/2+j)} b^{2j} \hat{\Delta}^j \phi(\vec{r}) \ ,
    \label{eq:pizzetti} 
\end{equation}
where $\Gamma(...)$ is the gamma function and $\hat{\Delta}$ is the Laplacian operator. EPD identity for this spherical mean -- at least, away from singularities -- is given by \cite{john2004plane}:
\begin{equation}
    \frac{\partial^2}{\partial b^2} \Phi(\vec{r},b) + \frac{(n-1)}{b} \frac{\partial}{\partial b} \Phi(\vec{r},b) = \hat{\Delta}_{\vec{r}} \Phi(\vec{r},b) \ ,
\label{eq:EPD}
\end{equation}
in which the Laplacian operator $\hat{\Delta}_{\vec{r}} $ is with respect to the spatial variable $\vec{r}$.

\subsection{The Requirement for Monopolar Shape-Preserving Property \label{sec:spp_mono_req}}

To satisfy the \textit{shape-preserving property}, the force generated by the potential $\Phi(\vec{r},b)$, i.e.
$$ \vec{F}(\vec{r},b) = - \vec{\nabla}_{\vec{r}} \Phi(\vec{r},b) \ , $$
must obey
\begin{equation}
    \vec{F}(\vec{r},b)=M(b) \vec{F}\left(g(b)\vec{r},0\right) \ ,
    \label{eq:same_shape_force}
\end{equation}
where $M(b)$ is the \textit{strength renormalization} factor and $g(b)$ is the radial \textit{rescaling factor} to be determined. The \textit{spherical property} corresponds to the special case $g(b)=1$, while the \textit{shell theorem} further requires $M(b)=1$.

We show in Appendix \ref{app:spherical_symmetry} that the force field satisfying Eq. \eqref{eq:same_shape_force} and its corresponding potential must be purely radial:
\begin{equation}
    \Phi(\vec{r},b) = \Phi(r,b) \ \ \text{and} \ \ F(r,b) = M(b) f(g(b)r) \ ,
    \label{eq:from_f_to_F}
\end{equation}
where $f(r)$ is the radial force field created by a point-source, i.e. $f(r)=-\phi'(r)$. Here, we use the notation $'$ for taking the derivative with respect to the only input argument. We can then apply the EPD identity to constrain the possible form of the force by differentiating both sides of Eq. \eqref{eq:EPD} with respect to $r$:
\begin{equation}
\frac{\partial^2}{\partial b^2}F(r,b)
+
\frac{n-1}{b} \frac{\partial}{\partial b} F(r,b)
=
\hat{L}_1 F(r,b) \ ,
\label{eq:force_EPD}
\end{equation}
where the operator is given by:
\begin{equation}
    \hat{L}_1 \equiv \frac{\partial^2}{\partial r^2} + \frac{n-1}{r} \frac{\partial}{\partial r} - \frac{n-1}{r^2} \ .
    \label{eq:L1}
\end{equation}

\subsection{Classifying all Monopolar Shape-Preserving Interactions \label{sec:mono_class}}

Define the operator \begin{equation}
    \hat{\Xi} = r \frac{d}{dr} = \frac{d}{d \ln(r)} \ ,
    \label{def:Xi}
\end{equation}
then
$$\hat{\Xi} f(r) = rf'(r) \ \ \text{and} \ \ \hat{\Xi}^2 f(r) = r^2 f''(r) + rf'(r) \ . $$ 
Following Section \ref{sec:spp_mono_req}, it can be shown that the radial force field $f(r)$ must satisfy
\begin{equation}
    \hat{L}_1 f(r) = \alpha f(r) + \beta \hat{\Xi}f(r) \ \ \text{for constants} \ \ \alpha, \beta
    \label{eq:perturb}
\end{equation} 
and
\begin{equation}
    g'^2(b) \hat{\Xi}^2 f(r) = P_1(b) f(r) + P_2(b) \hat{\Xi} f(r) \ ,
    \label{eq:ff_ode}
\end{equation} 
where $P_1(b)$ and $P_2(b)$ are functions of the shell radius $b$, whose derivation and explicit forms are provided in Appendix \ref{app:ff_ode}.
\begin{itemize}
    \item Let us first consider Eq. \eqref{eq:ff_ode}; this equation admits two possibilities:
\begin{itemize}
    \item \underline{$g(b)=1$:} This is precisely the requirement that $f(r)$ satisfy the spherical property, which has already been studied in \cite{tse2026generalizing}. In this case, \eqref{eq:ff_ode} reduces to
    $$ 0= P_1(b) f(r) + P_2(b) \hat{\Xi} f(r) $$
    where 
    $$ P_1(b)=\alpha - \frac{M''(b)}{M(b)} - \frac{n-1}{b} \frac{M'(b)}{M(b)} \ , \ P_2(b)=0 \ . $$
    For a nontrivial function $f(r)$, this implies $P_1(b)$ = 0, consistent with the result obtained in \cite{tse2026generalizing}. This condition can be solved directly, and we refer to Appendix \ref{app:monopolar} for a complete classification of the corresponding interaction potentials.
    \item \underline{$g(b)$ varies:} Eq. \eqref{eq:ff_ode}  becomes
    \begin{equation}
    \hat{\Xi}^2 f(r) = Q_1(b) f(r) + Q_2(b) \hat{\Xi} f(r) \ \ \text{where} \ \ Q_1(b)=\frac{P_1(b)}{g'^2(b)} \ , \ Q_2(b)=\frac{P_2(b)}{g'^2(b)} \ .
    \label{eq:gb_neq_1}
\end{equation} 
For non-trivial force field $f(r)$ that satisfy this equation for every shell radius $b$, this implies $Q_1(b)$ and $Q_2(b)$ are constants (explained in Appendix \ref{app:extra_case}), thus:
$$ \hat{\Xi}^2 f(r) = Q_1 f(r) + Q_2 \hat{\Xi} f(r) \ . $$
This ordinary differential equation (ODE) can be solved to get $f(r)$ be a linear combination of two power-laws, i.e.
\begin{equation}
    f(r) = U_- r^{\sigma_-} + U_+ r^{\sigma_+} \ ,
    \label{eq:2power}
\end{equation}
where $U_\mp$ are constants and
$$ \sigma_\mp = \frac12 \left[ Q_2 \mp \left( Q_2^2 + 4 Q_1 \right)^{1/2}\right] \ . $$
We note that the above line of arguments actually does not exhaust all cases. The remaining possibilities are treated in Appendix \ref{app:extra_case}, where we show that they reduce to known solutions for \textit{monopolar spherical property}.

\end{itemize}

\item Then, taking Eq. \eqref{eq:2power}, we proceed by considering the conditions under which it also satisfies \eqref{eq:perturb}; we find two possibilities:
\begin{itemize}
    \item \underline{$f(r)$ is regular at $r=0$:} The most general solution of this type is
    $$ f(r) = U_- r + U_+ r^3 \ , $$
    which corresponds to the interaction potential
    \begin{equation}
        \phi(r) = U_1 r^2 + U_2 r^4
        \label{eq:pot_reg}
    \end{equation}
    for constants $U_1=-U_-/2$ and $U_2=-U_+/4$. This is not included in the set of potentials possessing the \textit{monopolar spherical property} (see Appendix \ref{app:monopolar}). We can use the Pizzetti expansion in Eq. \eqref{eq:pizzetti} to obtain $\Phi(r,b)$ from Eq. \eqref{eq:pot_reg}:
    $$ \Phi(r,b) = \left[ U_1 + 2\left(1 + \frac2n \right) U_2 b^2\right] r^2 + U_2 r^4 + (U_1 + U_2 b^2) b^2 \ , $$
    then get the force field created by a shell of radius $b$:
    $$ F(r,b) = -2\left[ U_1 + 2\left(1 + \frac2n \right) U_2 b^2\right] r - 4U_2 r^3 $$
    to identify that the \textit{strength renormalization} and \textit{radial rescaling} factors are
    \begin{equation}
        M(b) = H^{3/2}(b) \ , \ g(b) =  H^{-1/2}(b) \ \ \text{where} \ \ H(b) = 1 + 2\left( 1 + \frac2n \right)  \frac{U_2}{U_1} b^2 \ .
        \label{eq:pot_reg_factor}
    \end{equation}
    \item \underline{$f(r)$ is singular at $r=0$:} The most general solution of this type is
    $$ f(r) = U_- r^{1-n} + U_+ r^{3-n} \ , $$
    which corresponds to the interaction potential
    \begin{equation}
\phi(r)=
\begin{cases}
\displaystyle{U_1 \ln(r) + U_2 r^2}
& \text{for} \ \ n=2 \ ,
\\[16pt]
\displaystyle{\frac{U_1}{r^2} + U_2 \ln(r)}
& \text{for} \ \ n=4 \ ,
\\[16pt]
\displaystyle{U_1r^{2-n}
    +
    U_2r^{4-n}}
& \text{for} \ \ n\neq 2,4 \ ,
\end{cases}
\label{eq:pot_sing}
\end{equation}
in which the constants can be calculated directly from $n$ and $U_\pm$. For dimension $n=2$, this potential is already included in the set of potentials possessing the \textit{monopolar shell theorem} (see Appendix \ref{app:monopolar}) and therefore does not offer anything new. We further consider $n \neq 2$ and apply the Pizzetti expansion in Eq. \eqref{eq:pizzetti} to obtain $\Phi(r,b)$ from Eq. \eqref{eq:pot_sing}:
\begin{equation*}
\Phi(r,b)=
\begin{cases}
\displaystyle{\frac{4U_1 + U_2 b^2}{4r^2} + U_2 \ln(r)}
& \text{for} \ \ n=4 \ ,
\\[16pt]
\displaystyle{\left[
        U_1+\left(\frac{4}{n}-1\right)U_2b^2
    \right]r^{2-n}
    +
    U_2r^{4-n}}
& \text{for} \ \ n\neq 2,4 \ ,
\end{cases}
\end{equation*}
then get the force field created by a shell of radius $b$:
\begin{equation*}
F(r,b)=
\begin{cases}
\displaystyle{\frac{4U_1 + U_2 b^2}{2r^3} - \frac{U_2}{r}}
& \text{for} \ \ n=4 
\\[16pt]
\displaystyle{(n-2)
    \left[
        U_1+\left( \frac{4}{n}-1\right)U_2b^2
    \right]r^{1-n}
    +(n-4)U_2r^{3-n}}
& \text{for} \ \ n\neq 2,4
\end{cases}
\end{equation*}
to identify that the \textit{strength renormalization} and \textit{radial rescaling} factors are
\begin{equation*}
\begin{cases}
\displaystyle
M(b)=g(b)=\left(1+\frac{U_2}{4U_1}b^2\right)^{-1/2}
& \text{for } n=4,
\\[16pt]
\displaystyle
M(b)=H^{(3-n)/2}(b) \ , \ g(b)=H^{-1/2}(b) 
& \text{for } n\neq2,4 \ .
\\[-2pt]
\displaystyle \ \ \ \ \ \ \ \ \ \ \ \ \ \ \ \
\text{where} \ \
H(b)=1+\left(\frac{4}{n}-1\right) \frac{U_2}{U_1}b^2 \ .
&
\end{cases}
\end{equation*}
\end{itemize}
\end{itemize}
We demonstrate monopolar shape preservation with an example in $n=2$-dimensional space in Fig. \ref{fig02}b, where the interacting potential is \textit{regular} and has the form given by Eq. \eqref{eq:pot_reg}:
$$
\phi(r) = \frac12 r^2 - \left( \frac{1-2^{-2/3}}8\right) r^4 \ .
$$
The resulting force field created by a \textit{monopolar} point-source is
$$
f(r) = -r + \left( \frac{1-2^{-2/3}}2\right) r^3,
$$
while that generated by a uniform circle of \textit{monopoles} with radius $b=1$ is
$$
F(r,b) = -2^{-2/3} r + \left( \frac{1-2^{-2/3}}2 \right) r^3 \ .
$$
These two force fields have the same shape in the sense of Eq. \eqref{eq:from_f_to_F}, as one can be mapped onto the other by a \textit{strength renormalization} factor $M(b)=1/2$ and a \textit{radial scaling} factor $g(b)=2^{1/3}$. Specifically, the mapping procedure
$$ \vec{f}(r) \ \longrightarrow \ M(b) \vec{f}(r) \ \longrightarrow \ M(b) \vec{f}(g(b)r) \ = \ \vec{F}(r,b) $$
is illustrated from left to right in Fig. \ref{fig02}b, shown within the spatial region $[-4,+4] \times [-4,+4]$.

We have now completely classified \textit{monopolar shape-preserving shell interaction} beyond \textit{spherical property}. What we do next, in Section \ref{sec:multipole}, is to generalize what we have found in this Section from \textit{monopolarity} to \textit{multipolarity}.

\section{Multipolar Shape-Preserving Shell Interactions \label{sec:multipole}}

We extend the analysis in Section \ref{sec:monopole} to uniformly \textit{aligned intrinsic multipoles} of arbitrary rank. Since spherical averaging preserves their angular structure, we can apply a version of the EPD identity to the radial multipolar potential profile. This yields the corresponding \textit{regular} and \textit{singular} families and shows how the classification depends on \textit{multipolar rank}.

\subsection{An Euler-Poisson-Darboux Identity for Multipolarity}

For any separable function of the form
$\rho(\vec{r})=u_\ell(r)\mathcal Y_\ell(\Omega)$, where
$\mathcal Y_\ell$ is a rank-$\ell$ \textit{hyperspherical harmonic}\footnote{A rank-$\ell$ \textit{hyperspherical harmonic} $\mathcal Y_\ell$ satisfies:
$$ \hat{\Delta}_{S^{n-1}}\mathcal Y_\ell
    =
    -\ell(\ell+n-2)\mathcal Y_\ell \ , $$
where $\hat{\Delta}_{S^{n-1}}$ is the Laplace–Beltrami operator on the unit $(n-1)$-hypersphere $S^{n-1}$.
In general, 
$$\mathcal Y_\ell(\Omega) = \sum_{\mathbf m}
    c_{\mathbf m}
    Y_{\ell\mathbf m}(\Omega) \ ,
$$
where $\mathbf m$ collectively labels the components needed to distinguish the independent hyperspherical harmonics at fixed $\ell$, and $c_{\ell \mathbf m}$ are the corresponding expansion coefficients. 
} \cite{atkinson2012spherical,ehrentraut1998symmetric}, the
Laplacian $\hat{\Delta}$ separates as:
\begin{equation}
    \hat{\Delta} \rho(\vec{r})
    =
    \left[
        \hat{L}_{\ell} u_\ell(r)
    \right]\mathcal Y_\ell(\Omega) \ , 
    \label{eq:Lap_sep}
\end{equation}
in which the operator $\hat{L}_{\ell}$ acts only on the radial distance coordinate $r$:
\begin{equation}
    \hat{L}_{\ell}
    \equiv
    \frac{\partial^2}{\partial r^2}
    +\frac{(n-1)}{r}\frac{\partial}{\partial r}
    -
    \frac{\ell(\ell+n-2)}{r^2} \ ;
\label{eq:curved_radial_operator_flat}
\end{equation}
note that for $\ell=1$ this operator reduces to $\hat{L}_1$ in Eq.\eqref{eq:L1}. The interaction potential created at position $\vec{r}$ by a rank-$\ell$ point multipole located at the origin $O$ is given by a rank-$\ell$ angular function:
\begin{equation}
    \phi_\ell(\vec{r})
    =
    \psi_\ell(r) \mathcal Y_\ell(\Omega) \ ,
\label{eq:curved_point_multipole}
\end{equation}
in which $\mathcal Y_\ell(\Omega)$ specifies the angular structure of the multipole and the radial function $\psi_\ell(r)$ is determined by the underlying radial potential $\phi(r)$:
\begin{equation} \psi_\ell(r) = \hat{D}_{\ell} \phi(r) \ , \ \ \text{where} \ \  \hat{D}_{\ell} \equiv r^\ell \left( \frac{1}{r}\frac{d}{dr} \right)^\ell \ . \label{eq:multipole_operator} \end{equation}
We provide a derivation of this using Hobson's differential formula \cite{nomura2018proof} in Appendix \ref{app:covariant_multipole}. 

\ \ 

Next, let us smear the point multipole into infinitely many identical microscopic multipoles uniformly distributed over an $(n-1)$-dimensional hypersphere $S^{n-1}(b)$ of radius $b$ centered at the original location $O$. In flat space, the multipoles can be naturally aligned by pure translation, so that each microscopic multipole has the same orientation as the original multipole at $O$. The potential generated by this hyperspherical shell of aligned multipoles is therefore given by the normalized spherical mean:
\begin{equation}
    \Phi_\ell(\vec{r},b)
    =\frac{ \oint_{S^{n-1}(b)} d\sigma(b\vec{e}_s) \ \phi_\ell \left(  \vec{r}-b\vec{e}_s  \right)}{\text{Vol}\left[ S^{n-1}(b)\right]} \ ,
\label{eq:spherical_mean_ell}
\end{equation}
where $S^{n-1}(b)$ is the $(n-1)$-dimensional spherical surface of radius $b$, $\mathrm{Vol}[S^{n-1}(b)]$ is its total surface volume, $\vec{e}_s$ is a unit direction vector on the sphere, and $d\sigma(b \vec{e}_s)$ is the surface-area element. Since the spherical mean is a rotationally invariant linear operator, it preserves the orientation structure associated with any rank-$\ell$ angular function such as Eq. \eqref{eq:curved_point_multipole}:
\begin{equation}
    \Phi_\ell(\vec{r},b) \equiv \Psi_\ell (r,b) \mathcal{Y}_\ell(\Omega) \ ;
    \label{eq:get_F}
\end{equation}
as a demonstration, we derive this for $n=3$ in Appendix \ref{app:angular_kept}. As a consequence of the Pizzetti expansion \cite{sabelfeld1997spherical} and the separability of the Laplacian operator i.e. \eqref{eq:Lap_sep}, we can calculate $\Psi_\ell(r,b)$ from the derivatives of $\psi_\ell(r)$ as: 
\begin{equation}
    \Psi_\ell(r,b) = \sum^\infty_{j=0} \frac{\Gamma(n/2)}{4^j j! \Gamma(n/2+j)} b^{2j} \hat{L}_\ell^j \psi_\ell(r) \ .
    \label{eq:Pizzetti_ell}
\end{equation}
Furthermore, the EPD identity for the spherical mean Eq. \eqref{eq:spherical_mean_ell} -- at least, away from singularities -- is given by the following PDE \cite{john2004plane}:
\begin{equation}
    \partial_b^2 \Phi_\ell(\vec{r},b) + \frac{(n-1)}{b} \partial_b \Phi_\ell(\vec{r},b) = \hat{\Delta}_{\vec{r}} \Phi_\ell(\vec{r},b) \ .
\label{eq:EPD_ell}
\end{equation}
Using Eq. \eqref{eq:get_F}, we arrive at a PDE constraint for the rank-$\ell$ coefficient $\Psi_\ell(r,b)$:
\begin{equation}
    \partial_b^2  \Psi_\ell (r,b) + \frac{(n-1)}{b} \partial_b  \Psi_\ell (r,b) = \hat{L}_\ell \Psi_\ell (r,b) \ .
\label{eq:EPD_rank_ell}
\end{equation}

\subsection{The Requirement for Multipolar Shape-Preserving Property \label{sec:spp_multi_req}}

In direct analogy with Section \ref{sec:spp_mono_req}, the force fields created by a \textit{multipolar shell} of radius $b$ can be calculated with:
\begin{equation}
    \vec F_{\ell}(\vec r,b)
    \equiv-\vec\nabla_{\vec r}\Phi_{\ell}(\vec r,b) \ ,
    \label{eq:multipole_force_pot}
\end{equation}
and in the \textit{point multipole} limit $b \rightarrow 0$ we have:
$$\Phi_\ell (\vec{r},0) = \phi_\ell(r) \ , \ \vec{f}_\ell(\vec{r}) \equiv -\vec{\nabla} \phi_\ell (\vec{r}) =  -\vec{\nabla}_{\vec{r}} \Phi_{\ell}(\vec r,0) = \vec{F}_\ell (\vec{r},0) \ . $$
The general \text{rank-$\ell$ shape-preserving} condition is
\begin{equation}
    \vec F_{\ell}(\vec r,b)
    =M_{\ell}(b)\,
      \vec f_{\ell}\!\left(g_{\ell}(b)\vec r\right) \ ,
    \label{eq:multipolar-shape-preserving-condition}
\end{equation}
where $M_{\ell}(b)$ is the \textit{strength renormalization} factor and
$g_{\ell}(b)$ is the \textit{radial rescaling} factor. We can further show that:
\begin{equation}
    \Psi_{\ell}(r,b)
    =N_{\ell}(b)\psi_{\ell} \left(g_{\ell}(b)r\right) \ \ \text{where} \ \ N_\ell(b) \equiv \frac{M_\ell(b)}{g_\ell(b)} \ , 
    \label{eq:multipolar_rel}
\end{equation}
with the derivation provided in Appendix \ref{app:multipolar_cond}. The \textit{multipolar spherical
property} is the special case $g_{\ell}(b)=1$, while the \textit{multipolar shell
theorem} further requires $M_{\ell}(b)=1$.

\subsection{Classifying all Multipolar Shape-Preserving Interactions}

Following Section \ref{sec:spp_multi_req}, it can be shown that the radial profile $\psi_\ell(r)$ must satisfy
\begin{equation}
    \hat L_{\ell}\psi_{\ell}(r)
    =\alpha_{\ell}\psi_{\ell}(r)
     +\beta_{\ell}\hat\Xi\psi_{\ell}(r) \ \ \text{for constants} \ \ \alpha_{\ell} , \beta_{\ell}
    \label{eq:multipolar_local_constraint}
\end{equation}
and
\begin{equation}
    {g'_{\ell}}^{2}(b)\hat\Xi^{2}\psi_{\ell}(r) = W_1(b)\psi_{\ell}(r)
    +W_2(b)\hat\Xi\psi_{\ell}(r) \ ,
    \label{eq:multipolar_master_equation}
\end{equation}
where $W_1(b)$ and $W_2(b)$ are functions of the shell radius $b$, whose derivation and explicit forms
can be found in Appendix \ref{app:derive_multipolar}.

\begin{itemize}
    \item Let us first consider Eq. \eqref{eq:multipolar_master_equation}; this equation admits two possibilities:
\begin{itemize}
    \item \underline{$g_{\ell}(b)=1$:} This corresponds to the \textit{multipolar spherical
    property}. Here, our main focus is the rank-$\ell>0$ \textit{multipolar shape-preserving interaction} beyond the \textit{spherical property}; the \textit{multipolar spherical property}, which to the best of our knowledge has not previously been studied, is treated separately in Appendix \ref{app:multipolar_sp}.

    \item \underline{$g_{\ell}(b)$ varies:} Eq. \eqref{eq:multipolar_master_equation}
    becomes
    \begin{equation}
        \widehat\Xi^{2}\psi_{\ell}(r)
        =X_1(b)\psi_{\ell}(r)
         +X_2(b)\widehat\Xi\psi_{\ell}(r) \ \ X_1(b)=\frac{W_1(b)}{{g_\ell'}^2(b)} \ , \ X_2(b)=\frac{W_2(b)}{{g'_\ell}^2(b)} \ .
        \label{eq:multipolar_Euler_b}
    \end{equation}
    Using arguments similar to those in Appendix \ref{app:extra_case}, for a nontrivial radial function $\psi_\ell (r)$ satisfying this equation for every shell radius $b$, $X_1(b)$ and $X_2(b)$ must be constants, yielding solutions (outside of those that have \textit{monopolar spherical property}) of the form:
    \begin{equation}
        \psi_{\ell}(r)
        =U_{-}r^{\sigma_{-}}+U_{+}r^{\sigma_{+}} \ ,
        \label{eq:multipolar_two_power_solution}
    \end{equation}
    where $U_\mp$ are constants and
    $$ \sigma_\mp = \frac12 \left[ X_2 \mp \left( X_2^2 + 4 X_1 \right)^{1/2}\right] \ . $$
    \end{itemize}
    \item Then, taking Eq. \eqref{eq:multipolar_two_power_solution}, we proceed by considering the conditions under which it also satisfies \eqref{eq:multipolar_local_constraint}; we find two possibilities, both of which can be written as:
    $$ \psi_{\ell}(r) = U_- r^k + U_+ r^{k+2} $$
    where $k \in \{ \ell , 2-n-\ell \}$. We then use Eq. \eqref{eq:Pizzetti_ell} to obtain:
    \begin{equation}
    \Psi_{\ell}(r,b)
    =U_{-} H_{\ell}(b)r^{k}+U_{+}r^{k+2} \ \ \text{where} \ \
    H_{\ell}(b)
    \equiv 1+\left( 1+\frac{2k}n\right) \frac{U_{+}}{U_{-}}b^{2} \ .
    \label{eq:unified_shell_coefficient}
\end{equation}
    Matching this result with \eqref{eq:multipolar_rel} gives us the identification:
    \begin{equation}
    M_{\ell}(b)=H_{\ell}^{(k+1)/2}(b) \ , \ g_{\ell}(b)=H_{\ell}^{-1/2}(b) \ . \label{eq:unified_multipolar_factors}
\end{equation}
    \begin{itemize}
        \item \underline{$k=\ell$:} The corresponding radial potential $\phi(r)$, which can be found by solving Eq. \eqref{eq:multipole_operator}, is regular at $r=0$:
        \begin{equation}
            \phi(r) = U_1 r^{2\ell}+U_2 r^{2(\ell+1)}
            +\sum_{j=0}^{\ell-1}C_jr^{2j} \ ,
            \label{eq:multi_reg}
        \end{equation}
        where $U_1$ and $U_2$ are constants that can be calculated from $U_\pm$ and $\ell$. The $C$-values can be any arbitrary constants. The \textit{strength renormalization} and the \textit{radial rescaling} factors are:
        \begin{equation}
        \begin{split}
        M_\ell(b) = H_\ell^{(\ell+1)/2}(b) \ , & \ g_\ell(b) = H_\ell^{-1/2}(b) 
        \\
        \text{where} \ \ & H_\ell(b)=1+(\ell+1)\left(1+\frac{2\ell}{n} \right)
             \frac{U_2}{U_1}b^{2}
        \ .
        \end{split}
    \label{eq:reg_shape_factors}
    \end{equation}
    \item \underline{$k=2-n-\ell$:} Solving  Eq. \eqref{eq:multipole_operator} gives us singular radial potential $\phi(r)$ at $r=0$:
        \begin{equation}
            \phi(r)=
    \begin{cases}
        U_1 \ln(r)+U_2r^{2} 
        & \text{for} \ \ n=2 \ ,\ \ell=1 \ , \\[4pt]
        U_1 \ln(r) + U_2 r^{2} \ln(r) 
        & \text{for} \ \ n=2 \ ,\ \ell\geq2 \ ,\\[4pt]
        \displaystyle\frac{U_1}{r^{2}}+U_2 \ln(r)
        & \text{for} \ \ n=4 \ , \\[10pt]
        U_1 r^{2-n}+U_2 r^{4-n},
        &  \text{for} \ \ n\neq2 ,4 \ ,
    \end{cases}
    \ \ \ \ +\sum_{j=0}^{\ell-1}C_jr^{2j} \ .
            \label{eq:multi_sing}
        \end{equation}
        where $U_1$ and $U_2$ are constants that can be determined from $U_\pm$, $n$, and $\ell$. The $C$-values can be any arbitrary constants. The corresponding \textit{strength renormalization} and the \textit{radial rescaling} factors are:
        \begin{equation}
            M_{\ell}(b)=H_{\ell}^{(3-n-\ell)/2}(b) \ , \ g_{\ell}(b)=H_{\ell}^{-1/2}(b)
            \ ,
            \label{eq:sing_shape_factors}
        \end{equation}
where
\begin{equation}
    H_{\ell}(b)=
    \begin{cases}
        1
        & \text{for} \ \ n=2 \ ,\ \ell=1 \ , \\[4pt]
        1 + \displaystyle{\frac{U_2}{U_1}} b^2 
        & \text{for} \ \ n=2 \ ,\ \ell\geq2 \ , \\[4pt]
        1+ \displaystyle{\frac{U_2}{4 U_1}} b^2
        & \text{for} \ \ n=4 \ , \\[10pt]
        1+\displaystyle{\left(\frac4n -1\right)  \frac{U_2}{U_1}} b^2 
        &  \text{for} \ \ n\neq2 , 4 \ .
    \end{cases}
    \label{eq:singular_H_factor}
\end{equation}
For $n=2$ and $\ell=1$,
$H_{\ell}=1$ and the result reduces to the \textit{multipolar shell theorem} rather
than producing something new (see Appendix \ref{app:multipolar_sp}).
\end{itemize}
\end{itemize}

\begin{figure*}[thbp!]
    \centering
\includegraphics[width=\linewidth]{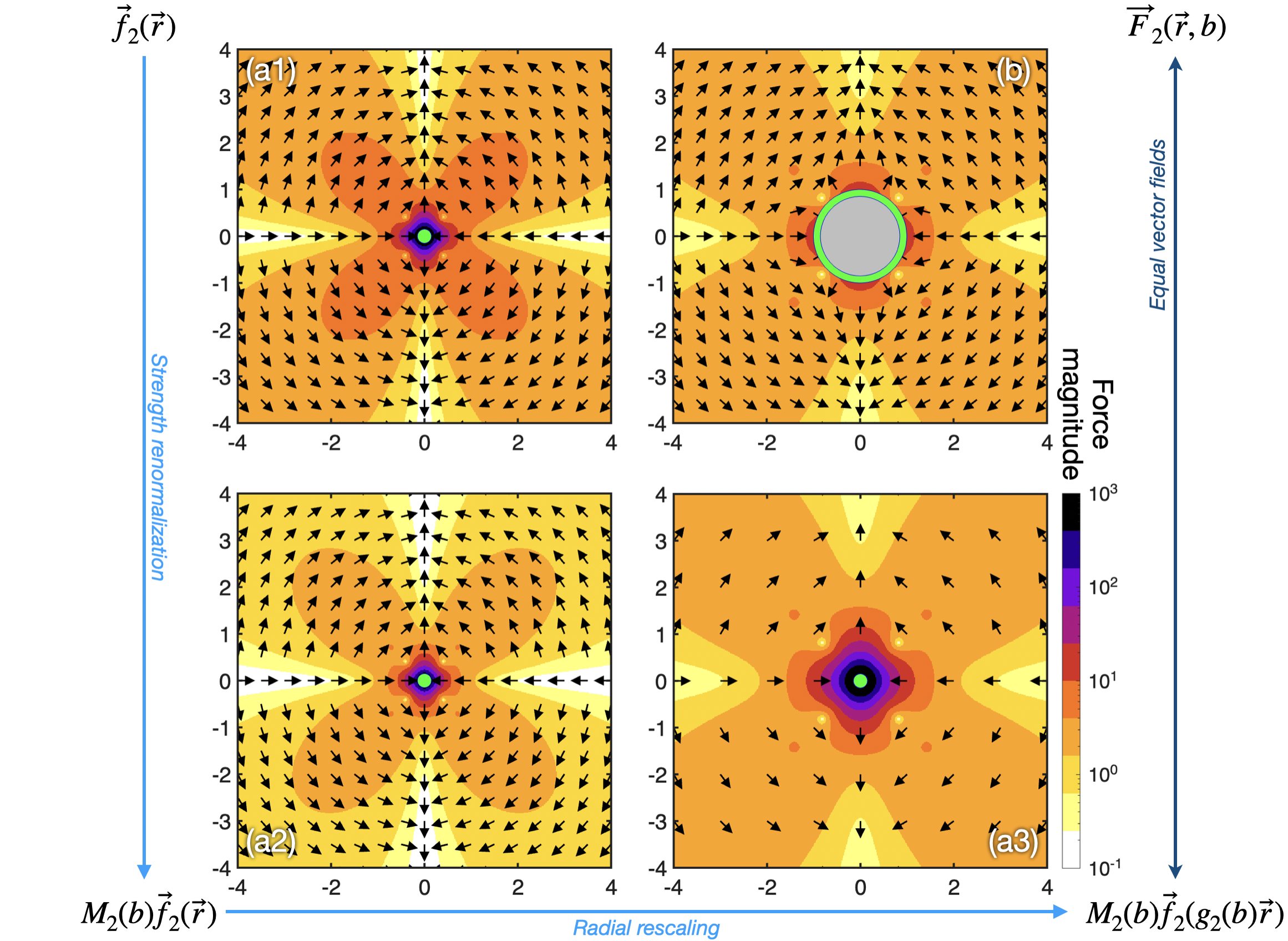}
    \caption{\textbf{Multipolar shape-preservation.} We show the mapping of the multipolar point-source force field to that of the uniform circle shell through \textit{strength renormalization} and \textit{radial rescaling}: \textbf{(a1)} the force field created by the original point-source, \textbf{(a2)} the \textit{strength renormalized} force field, \textbf{(a3)} the \textit{strength renormalized} and \textit{radial rescaled} force field, and \textbf{(a3)} the force field created if the point-source got smeared uniformly on a circle of radius $b=1$. The green dots indicate the point-sources and the hollow green circle represents the spherical shell source, while the \textit{fire}-colormap represents the force magnitude. Arrows indicate the force direction, which is purely radial.}
\label{fig03}
\end{figure*}

It is important to note that these \textit{multipolar shape-preserving interactions} admit an additional polynomial potential between rank-$(\ell-1)$ and rank-$\ell$, i.e. a contribution $
\propto r^{2(\ell-1)}$, which, despite being nontrivial, generates no rank-$\ell$ multipole field as follows from Eq. \eqref{eq:multipole_operator}. In other words, these potentials are \textit{invisible} to the rank-$\ell$ multipole field, i.e. the \textit{kernel} of the \textit{multipolar derivative operator} $\hat{D}_\ell$:
\begin{equation}
    \text{ker} \hat{D}_\ell = \text{span} \Big\{ 1,r^2,r^4,...,r^{2(\ell - 1)}\Big\}\ .
    \label{eq:kernel}
\end{equation}
Modulo this kernel, the \textit{multipolar} classification differs from the \textit{monopolar} one (see Section \ref{sec:mono_class}). The \textit{regular branch} produces the rank-$\ell$ visible potentials
$$U_1 r^{2\ell} + U_2 r^{2(\ell+1)}$$
which are not \textit{monopolar shape-preserving interaction} for $\ell\geq2$ in any dimension. The \textit{singular branch}, on the other hand, yields potentials identical to those of the \textit{monopolar shape-preserving interaction}, except for the additional family
$$U_1 \ln(r) + U_2 r^2 \ln(r)$$
that appears only at $n=2$ for $\ell\geq2$. We demonstrate multipolar shape preserving property of a interacting potential of this form, in $n=2$-dimensional space for $\ell=2$:
$$ \phi(r) = \ln(r) + 3 r^2 \ln(r) \ . $$
For an aligned quadrupolar point-source, up to a spatial rotation, the corresponding multipolar potential is given by
$$\phi_2(\vec{r})
=
\psi_2(r)\cos(2\theta)
=
\left(-\frac{2}{r^2}+6\right)\cos(2\theta)  \ . $$
The resulting force field created by a \textit{multipolar} point-source is
$$ \vec{f}_2(\vec{r}) = -\frac{4}{r^3} \cos(2\theta) \vec{e}_r + \left( -\frac{4}{r^3} + \frac{12}{r}\right) \sin(2\theta) \vec{e}_\theta \ , $$
where $\vec{e}_r$ is the radial unit vector and $\vec{e}_\theta$ is the tangential unit vector. On the other hand, the force field created by a uniform circle of \textit{multipoles} with radius $b=1$ is
$$ \vec{F}_2(\vec{r},b) = -\frac{16}{r^3} \cos(2\theta) \vec{e}_r + \left( -\frac{16}{r^3} + \frac{12}{r}\right) \sin(2\theta) \vec{e}_\theta \ . $$
These two force fields have the same shape in the sense of Eq. \eqref{eq:multipolar-shape-preserving-condition}, as one can be mapped
onto the other by a strength \textit{renormalization factor} $M_2 (b) = 1/2$ and a \textit{radial scaling} factor
$g_2(b)=1/2$. We illustrate this mapping procedure in Fig. \ref{fig03} (from subfigures a1, a2, a3, to b, sequentially):
$$ \vec{f}_2(\vec{r}) \ \longrightarrow \ M_2(b) \vec{f}_2(\vec{r}) \ \longrightarrow \ M_2(b) \vec{f}(g_2(b)\vec{r}) \ = \ \vec{F}_2(\vec{r},b) \ , $$
shown within the spatial region $[-4,+4] \times [-4,+4]$. 

\section{Discussion \& Conclusion}

We have classified the radial interactions for which the exterior field
of a uniform spherical shell of monopoles is a
\textit{strength renormalized} and \textit{radial rescaled} copy of
the corresponding point-source field. Beyond the 
\textit{spherical property} \cite{tse2026generalizing}, which includes Gurzadyan's cosmological
theorem \cite{gurzadyan1985cosmological} as a special case, we identify two additional genuinely
dilational families: one \textit{regular} and one \textit{singular}. We further extend the shape-preserving condition to uniformly \textit{aligned intrinsic multipoles} of arbitrary rank. In this setting, spherical averaging preserves the multipole rank and angular orientation, so the classification is determined entirely by the radial multipolar profile. Similar to the monopolar case, two genuinely dilational branches arise. The \textit{regular} branch depends explicitly on the multipole rank and shifts toward progressively higher even powers of the underlying radial interaction as the rank increases. By contrast, the \textit{singular} branch is largely inherited from the monopolar classification, apart from logarithmic forms that arise in the critical dimensions. We summarize previously known results and our findings in this work in Table \ref{tab:known_and_new_results}.

\begin{longtblr}[
  caption={
    Summary of related sphere-to-point reductions and the
    shape-preserving classification discovered in this work.
  },
  label={tab:known_and_new_results}
]{
  width=0.9\textwidth,
  colspec={
    X[0.65,l]
    X[1.25,l]
    X[1.55,l]
  },
  rowhead=1,
  row{1}={font=\bfseries},
  cells={mode=text,valign=t,font=\small},
  colsep=4pt,
  rowsep=3pt,
}

\toprule

Setting
&
Earlier results
&
Results found here
\\

\midrule

Newtonian shell theorem
&
Exact exterior equivalence between a uniform shell and a central point
source for the Newtonian interaction
\cite{newton1833philosophiae,sneddon1949camb,kilmister1974newton}.
&
Recovered as a special case when $M(b)=g(b)=1$ in Eq. \eqref{eq:from_f_to_F}.
\\

Monopolar pherical property
&
The shell field may differ from the point-source field by a
shell-radius-dependent strength factor
\cite{chapman1983gravity,barnes1984gazette,tse2026generalizing}.
&
Recovered as the whole branch $g(b)=1$ in Eq. \eqref{eq:from_f_to_F}.
\\

Selected multipolar shell theorem
&
Exact reductions are known for uniformly magnetized spheres and
inverse-distance shells of aligned multipoles
\cite{edwards2017interactions,ji2021potential}.
&
Recovered as special cases with
$M_\ell(b)=g_\ell(b)=1$ in Eq. \eqref{eq:multipolar-shape-preserving-condition}.

\\

Multipolar spherical property
&
None
&
Recovered as the whole branch $g_\ell (b)=1$ in Eq. \eqref{eq:multipolar-shape-preserving-condition}; details can be found in Appendix \ref{app:multipolar_sp}.

\\

Monopolar shape-preservation
&
None
&
Allowing $g(b)$ to vary in Eq. \eqref{eq:from_f_to_F} requires radial interaction to be of the form in Eq. \eqref{eq:pot_reg} or Eq. \eqref{eq:pot_sing}.

\\

Multipolar shape-preservation
&
None
&
Allowing $g_\ell(b)$ to vary in Eq. \eqref{eq:multipolar-shape-preserving-condition} requires radial interaction to be of the form in Eq. \eqref{eq:multi_reg} or Eq. \eqref{eq:multi_sing}.

\\

\bottomrule

\end{longtblr}

There are several directions we could explore from here. One is to generalize the construction to \textit{curved spaces}, where \textit{defining alignment} for multipoles becomes more difficult because \textit{parallel transport} depends on the path. Another is to relax the shape-preserving condition and ask whether the potential can instead \textit{remain within a finite-dimensional class} of linear combinations of known potentials. A further direction is to consider \textit{dynamical radiative} settings, asking when the exterior field of an extended source can be represented by an effective point monopole/multipole after accounting for \textit{finite propagation speed} and \textit{frequency-dependent renormalization}. Finally, it would be natural to move \textit{beyond spherical shells} and study extended structures with more general shapes and nontrivial topologies.

\textbf{AI Transparency Statement:} The authors used generative AI tools for copy-editing, exposition, and formatting. The scientific ideas (as a follow-up to \cite{tse2026generalizing}), numerical investigations, theoretical derivations, and conclusions were developed and verified by the authors, who take full responsibility for the manuscript.

\bibliographystyle{apsrev4-2}
\bibliography{arxiv_references}

@article{gurzadyan1985cosmological,
  title={The cosmological constant in the McCrea-Milne cosmological scheme},
  author={Gurzadyan, Vahagn G},
  journal={The Observatory},
  volume={105},
  pages={42--43},
  year={1985}
}

@article{reed2022note,
  title={A note on Newton's shell-point equivalency theorem},
  author={Reed, B Cameron},
  journal={American Journal of Physics},
  volume={90},
  number={5},
  pages={394--396},
  year={2022},
  publisher={AIP Publishing}
}

@article{Carimalo2023,
  author    = {Christian Carimalo},
  title     = {Comment on the article ``A note on Newton's shell-point equivalency theorem'' [Am. J. Phys. 90, 394 (2022)]},
  journal   = {American Journal of Physics},
  volume    = {91},
  number    = {12},
  pages     = {1031},
  year      = {2023},
  publisher = {American Association of Physics Teachers},
  doi       = {10.1119/5.0153897},
  url       = {https://doi.org/10.1119/5.0153897}
}

@article{phan2025derivation,
  author       = {Phan, Trung V},
  title        = {On the derivation of the cosmological Gurzadyan's theorem},
  journal      = {Physics Education},
  year         = {2026},
  volume       = {61},
  number       = {1},
  pages        = {015003},
  doi          = {10.1088/1361-6552/ae153b},
  publisher    = {IOP Publishing},
}

@book{newton1833philosophiae,
  title={Philosophiae naturalis principia mathematica},
  author={Newton, Isaac},
  volume={1},
  year={1833},
  publisher={G. Brookman}
}

@article{sneddon1949camb,
  author    = {I. N. Sneddon and C. K. Thornhill},
  title     = {A Property of the Yukawa Potential},
  journal   = {Proceedings of the Cambridge Philosophical Society},
  volume    = {45},
  pages     = {318--320},
  year      = {1949}
}

@article{kilmister1974newton,
  title={Newton, the inverse square law and the direct distance law},
  author={Kilmister, CW},
  journal={The Mathematical Gazette},
  volume={58},
  number={403},
  pages={35--39},
  year={1974},
  publisher={Cambridge University Press}
}

@article{chapman1983gravity,
  author    = {C. J. Chapman},
  title     = {A Problem of Unusual Gravity},
  journal   = {The Mathematical Gazette},
  volume    = {67},
  number    = {439},
  pages     = {14--17},
  year      = {1983},
  publisher = {Mathematical Association},
  doi       = {10.2307/3617354}
}

@article{barnes1984gazette,
  author    = {A. Barnes and C. K. Keogh},
  title     = {Unusual Gravity},
  journal   = {The Mathematical Gazette},
  volume    = {68},
  pages     = {138--139},
  year      = {1984}
}

@article{carimalo2026note,
  title={A note on Gurzadyan theorem},
  author={Carimalo, Christian},
  journal={arXiv preprint arXiv:2603.25323},
  year={2026}
}

@book{sabelfeld1997spherical,
  title={Spherical means for PDEs},
  author={Sabelfeld, K Karl Karlovich and Shalimopva, IA},
  year={1997},
  publisher={VSP}
}

@book{john2004plane,
  title={Plane waves and spherical means applied to partial differential equations},
  author={John, Fritz},
  year={2004},
  publisher={Courier Corporation}
}

@article{tse2026generalizing,
  title={Generalizing the Shell Theorem to Constant Curvature Spaces in All Dimensions and Topologies},
  author={Tse, Ava K and Markowich, Olivia M and Phan, Trung V},
  journal={Gravitation and Cosmology},
  volume={32},
  number={3},
  pages={383--389},
  year={2026},
  publisher={Springer}
}

@article{edwards2017interactions,
  title={Interactions between uniformly magnetized spheres},
  author={Edwards, Boyd F and Riffe, D Mark and Ji, Jeong-Young and Booth, William A},
  journal={American Journal of Physics},
  volume={85},
  number={2},
  pages={130--134},
  year={2017},
  publisher={AIP Publishing}
}

@article{ji2021potential,
  title={Potential, field, and interactions of multipole spheres: Coated spherical magnets},
  author={Ji, Jeong-Young and Edwards, Boyd F and Spencer, J Andrew and Held, Eric D},
  journal={Journal of Magnetism and Magnetic Materials},
  volume={529},
  pages={167861},
  year={2021},
  publisher={Elsevier}
}

@book{atkinson2012spherical,
  title={Spherical harmonics and approximations on the unit sphere: an introduction. Lecture Notes in Mathematics},
  author={Atkinson, Kendall and Han, Weimin},
  volume={2044},
  year={2012},
  publisher={Springer Heidelberg}
}

@article{ehrentraut1998symmetric,
  title={On symmetric irreducible tensors in d-dimensions},
  author={Ehrentraut, Harald and Muschik, Wolfgang},
  journal={ARI-An International Journal for Physical and Engineering Sciences},
  volume={51},
  number={2},
  pages={149--159},
  year={1998},
  publisher={Springer}
}

@article{nomura2018proof,
  title={A proof of Hobson's formula with the Euler operator},
  author={Nomura, Takaaki},
  journal={Kyushu Journal of Mathematics},
  volume={72},
  number={2},
  pages={423--427},
  year={2018},
  publisher={Faculty of Mathematics, Kyushu University}
}

@book{varshalovich1988quantum,
  title={Quantum theory of angular momentum},
  author={Varshalovich, Dmitri{\u{\i}} Aleksandrovich and Moskalev, Anatol{\"{\i}} Nikolaevitch and Khersonskii, Valerij Kel'manov{\v{\i}}c},
  year={1988},
  publisher={World Scientific}
}

@book{hilbert1985methods,
  title={Methods of mathematical physics},
  author={Hilbert, David},
  year={1985},
  publisher={CUP Archive}
}

@book{cochran2022introduction,
  title={Introduction to sonar transducer design},
  author={Cochran, John C},
  year={2022},
  publisher={Wiley Online Library}
}

@article{kocher1978point,
  title={Point-multipole expansions for charge and current distributions},
  author={Kocher, Carl A},
  journal={American Journal of Physics},
  volume={46},
  number={5},
  pages={578--579},
  year={1978},
  publisher={AIP Publishing}
}

@article{ee2017combinatorics,
  title={Combinatorics in tensor-integral reduction},
  author={Ee, June-Haak and Jung, Dong-Won and Kim, U-Rae and Lee, Jungil},
  journal={European Journal of Physics},
  volume={38},
  number={2},
  pages={025801},
  year={2017},
  publisher={IOP Publishing}
}

@incollection{hall2013lie,
  title={Lie groups, Lie algebras, and representations},
  author={Hall, Brian C},
  booktitle={Quantum theory for mathematicians},
  pages={333--366},
  year={2013},
  publisher={Springer}
}

\appendix 

\section{Derivation of Eq. \eqref{eq:from_f_to_F} and Eq. \eqref{eq:force_EPD} \label{app:spherical_symmetry}}

Eq. \eqref{eq:same_shape_force} can be expressed as the requirement on the potential:
$$
\vec{\nabla}_{\vec{r}} \Phi(\vec r,b)= M(b) \vec{\nabla}_{\vec{a}} \Phi(\vec{a},0) \Big|_{\vec{a}=g(b)\vec{r}} = \frac{M(b)}{g(b)} \vec{\nabla}_{\vec{r}} \Phi(\vec{a},0) \Big|_{\vec{a} = g(b) \vec{r}} \ ,
$$
where we have used the chain rule. Because the limit $b\rightarrow 0$ corresponds to a point-source, we can make the identification $\Phi(\vec{r},0)=\phi(r)$ and rewrite the previous equation as:
$$ \vec{\nabla}_{\vec{r}} \Phi(\vec r,b) = \frac{M(b)}{g(b)} \vec{\nabla}_{\vec{a}} \phi(\vec{a}) \Big|_{\vec{a}=g(b) \vec{r}} \ . $$
Since both sides are gradients with respect to $\vec{r}$, integration gives
\begin{equation}
    \Phi(\vec{r},b) = \frac{M(b)}{g(b)} \phi\left(g(b)r \right) + C(b) \ ,
    \label{eq:Phi_from_phi}
\end{equation}
where $C(b)$ is a function of the shell radius $b$. Note that the right-hand side (RHS) depends only on the radial distance $r$ and is independent of the direction $\vec{r}/r$, the potential on the left-hand side (LHS) must be spherically symmetric with respect to the spatial variable $\vec{r}$, i.e. $\Phi(\vec{r},b) = \Phi(r,b)$ and thus:
$$\vec{F}(\vec{r},b) = F(r,b) \ \vec{r}/r \ \ \text{where} \ \ F(r,b) = -\frac{\partial}{\partial r} \Phi(r,b) \ . $$
We apply the EPD identity to constrain the possible form of the force, i.e. by differentiating both sides of Eq. \eqref{eq:EPD} with respect to $r$, and thus arrive directly at  Eq. \eqref{eq:force_EPD}. 

Define
$$-\vec{\nabla}_{\vec{r}} \phi(r) \equiv \vec{f}(\vec{r}) = f(r) \ \vec{r}/r \ \ \text{in which} \ \ f(r) =- \phi'(r) $$
and note that in the point-source limit $b \rightarrow 0$ we must have the identification
$$ \Phi(\vec{r},0) = \phi(r) \ , $$
then Eq. \eqref{eq:same_shape_force} can be rewritten as:
\begin{equation*}
    F(r,b) = M(b) f(g(b)r) \ ,
\end{equation*}
which is the expression in Eq. \eqref{eq:from_f_to_F}. Also, the boundary conditions at $b=0$ must be:
\begin{equation*}
M(0)=1 \ , \ g(0)=1 \ , \ C(0)= 0 \ .  
\end{equation*}
The integration Eq. \eqref{eq:full_potential} for $\Phi(\vec{r},b)$ is invariant under the transformation $b \rightarrow -b$, therefore we can also set:
\begin{equation*}
    M'(0)=0 \ , \ g'(0)=0 \ , \ C'(0)=0 \ .
\end{equation*}

\section{Derivation of Eq. \eqref{eq:perturb} and Eq. \eqref{eq:ff_ode} \label{app:ff_ode}}

Define:
$$a \equiv g(b)r \ , \ \mu(b) \equiv \frac{M'(b)}{M(b)} \ , \ \gamma(b) \equiv \frac{g'(b)}{g(b)} \ , $$
then after substituting Eq. \eqref{eq:from_f_to_F} into the EPD identity for the force Eq. \eqref{eq:force_EPD}, we can arrive at:
\begin{equation}
    O_1(b) f(a) + O_2(b) \hat{\Xi} f(a) + g'^2(b) \hat{\Xi}^2 f(a) = g^4(b) \hat{L}_1 f(a)  \ ,
    \label{eq:full_f_EPD_constraint}
\end{equation}
where
\begin{equation*}
    \begin{split}
        O_1(b) &= g^2(b) \left[ \mu'(b) + \frac{n-1}{b} \mu(b) + \mu^2(b) \right] \ , 
        \\
        O_2(b) &=g^2(b) \left\{ \gamma'(b) + \left[ 2\mu(b)  + \frac{n-1}{b} \right] \gamma(b) \right\} \ .
    \end{split}
\end{equation*}

If Eq. \eqref{eq:from_f_to_F} holds for all values of $b$, it must also hold for infinitesimal values of $b$ in a neighborhood of $b=0$. Consider the following Taylor expansions:
$$ M(b) \approx 1 + \frac12 M''(0) b^2 \ , \ g(b) \approx 1 + \frac12 g''(0) b^2 \ , $$
therefore:
$$M(b) f(g(b)r) \approx f(r) + \frac12 \left[ M''(0) f(r) + g''(0) \hat{\Xi} f(r) \right] b^2 \ . $$
Use the approximation
\begin{equation}
    F(r,b) \approx f(r) + \frac12 \left( \frac{1}{n} \hat{L}_1 f(r) \right) b^2 \ ,
    \label{eq:approx_force}
\end{equation}
which we derive in Appendix \ref{app:Taylor}, and matching the two sides of Eq. \eqref{eq:from_f_to_F} order-by-order in $b$, we find that the lowest nonzero contribution, which occurs at second order in $b$, should give us a matching ODE:
\begin{equation}
    \hat{L}_1 f(r) = \alpha f(r) + \beta \hat{\Xi}f(r) \ , 
    \label{eq:get_L1f}
\end{equation}
where $\alpha \equiv n M''(0)$ and $\beta \equiv n g''(0)$ are two constants -- this is Eq. \eqref{eq:perturb} in the main text. The derivation of this expression follows the \textit{perturbation method} explained in \cite{phan2025derivation}.

With the spatial variable $a$ relabeled as $r$, replace the LHS of Eq. \eqref{eq:full_f_EPD_constraint} with the RHS of Eq. \eqref{eq:get_L1f}, we arrive exactly at Eq. \eqref{eq:ff_ode}, in which the functions $P_1(b)$ and $P_2(b)$ are given explicitly with:
$$P_1(b) = \alpha g^4(b) - O_1(b) \ , \ P_2(b) = \beta g^4(b)- O_2(b) \ . $$

\section{Taylor Expansion of the Force \label{app:Taylor}}

The equivalent of Eq. \eqref{eq:full_potential}, for forces instead of potentials, is given by:
\begin{equation}
    \vec{F}(\vec{r},b) = \frac{ \oint_{S^{n-1}(b)} d\sigma(b \vec{e}_s) \ \vec{f} \left( \vec{r}-b\vec{e}_s \right)}{\text{Vol}\left[ S^{n-1}(b)\right]} \ ,
\label{eq:full_force}
\end{equation}
where in our physical setting we are dealing with the specific case:
$$\vec{f} \left( \vec{r}-b\vec{e}_s \right) = f \left(  | \vec{r}-b\vec{e}_s  | \right) \frac{\vec{r}-b\vec{e}_s }{| \vec{r}-b\vec{e}_s  |} \ . $$
The Taylor expansion for the force integrand, up to second-order in $b$, can be written as:
$$ \vec{f}(\vec{r}-b\vec{e}_s) = \vec{f}(\vec{r}) - \left[ \sum_{i=1}^n e_i \frac{\partial}{\partial r_i} \vec{f}(\vec{r}) \right] b + \frac12 \left[ \sum_{i,j=1}^n e_i e_j \frac{\partial}{\partial r_i} \frac{\partial}{\partial r_j} \vec{f}(\vec{r}) \right] b^2 \ , $$
where $i$, $j$ are the Cartesian indices ranging from $1$ to $n$ and $e_i$ is the $i$-th component of the unit-vector $\vec{e}_s$. Place this expression in Eq. \eqref{eq:full_force} and then use the following averages over the whole sphere \cite{ee2017combinatorics}:
$$ \frac{ \oint_{S^{n-1}(b)} d\sigma(b \vec{e}_s) \ e_i}{\text{Vol}\left[ S^{n-1}(b)\right]} = 0 \ , \ \frac{ \oint_{S^{n-1}(b)} d\sigma(b \vec{e}_s) \ e_i e_j}{\text{Vol}\left[ S^{n-1}(b)\right]} = \frac{\delta_{ij}}{n} \ \ \text{where $\delta_{ij}$ is the Kronecker-delta} \ \ , $$
we arrive at the approximation:
$$\vec{F}(\vec{r},b) \approx \vec{f}(\vec{r}) + \frac{1}{2} \left[ \frac{1}{n} \hat{\Delta} \vec{f}(\vec{r}) \right] b^2 \ . $$
For radial forces, i.e. $\vec{f}(\vec{r}) = f(r) \ \vec{r}/r$ and $\vec{F}(\vec{r},b) = F(r,b) \ \vec{r}/r$, we get:
$$ F(r,b) \approx f(r) + \frac{1}{2} \left[ \frac{1}{n} \hat{\Delta} f(r) \right] b^2 \ , $$
which is the same as Eq. \eqref{eq:approx_force} since
$$\hat{\Delta} f(r) = f''(r) + \frac{n-1}{r} f'(r) - \frac{n-1}{r^2} f(r) = \hat{L}_1 f(r) \ . $$

\section{Monopolar Spherical Property \label{app:monopolar}}

We have found in \cite{tse2026generalizing} that the \textit{general monopolar spherical property} is satisfied for $\lambda\neq0$, while the \textit{monopolar shell theorem} is obtained as the \textit{special} case $\lambda=0$; the full set of potentials satisfying these properties is given by:
\begin{itemize}
    \item \underline{$\lambda \neq 0$}: For $\lambda > 0$, the potential is a linear combination of the Bessel functions: 
    \begin{equation}
        \phi(r) = r^{-\alpha} \left[ A_1  J_{\alpha}(\kappa r) + A_2 Y_\alpha (\kappa r) \right]+ A_3 \ ,
        \label{eq:mono_neq0}
    \end{equation}
    where $\alpha = n/2 - 1$, $\kappa=\lambda^{1/2}$, and the $A$-values are arbitrary constants. The shell-\textit{radius-dependent rescaling factor} $M(b)$ can be found by solving the equation relating $M(b)$ with $\lambda$, i.e.
    $$ -\lambda M(b) = \partial_b^2 M(b) + \frac{(n-1)}{b} \partial_b M(b) \ , $$
    and consider the boundary conditions $M(0)=1$ and $M'(0)=0$, which gives:
    \begin{equation}
        M(b) = \Gamma\left(\frac{n}2 \right) \left( \frac{2}{\kappa b} \right)^\alpha J_\alpha (\kappa b) \ .
        \label{eq:mono_M}
    \end{equation}
    For $\lambda < 0$, it is obtained by analytic continuation.

\item \underline{$\lambda = 0$}: The potential is a linear combination between the particular solution and the fundamental solution of a constant source Poisson equation:
\begin{equation}
    \phi(r) = \begin{cases}
A_4 \ln (r)  + A_5 r^{2}+ A_6 
& \text{for} \ \ n=2 \ , \\[16pt]
\displaystyle{\frac{A_7}{a^{n-2}}} +A_8 r^{2}
+ A_9
& \text{for} \ \ n>2 \ ,
\end{cases} \ ,
    \label{eq:mono_0}
\end{equation}
where the $A$-values are arbitrary constants.
\end{itemize}

\section{Other Possible Solutions to Eq. \eqref{eq:gb_neq_1} \label{app:extra_case}}

Consider two different shell radii, e.g. $b_1$ and $b_2$, then Eq. \eqref{eq:gb_neq_1} gives:
\begin{equation*}
\begin{split}
    \hat{\Xi}^2 f(r) &= Q_1(b_1) f(r) + Q_2(b_1) \hat{\Xi} f(r)  \ ,
    \\
    \hat{\Xi}^2 f(r) &= Q_1(b_2) f(r) + Q_2(b_2) \hat{\Xi} f(r) \ , 
\end{split}
\end{equation*} 
Subtracting the first equation from the second one on both sides, we obtain:
$$\left[Q_1(b_2)-Q_1(b_1)\right]f(r)
+
\left[Q_2(b_2)-Q_2(b_1)\right]\hat{\Xi}f(r)
=0 \ . $$
\begin{itemize}
    \item For linearly independent $f(r)$ and $\hat{\Xi} f(r)$, this leads to $Q_1(b_1)=Q_1(b_2)$ and $Q_2(b_1)=Q_2(b_2)$ for any pair of $b_1$ and $b_2$, thus they are constants. We follow this direction in the main text.
    \item For linearly dependent $f(r)$ and $\hat{\Xi} f(r)$, $Q_1(b)$ and $Q_2(b)$ need not be constants; instead, their ratio must remain constant, so that linear dependence can be satisfied, i.e.
    $$\hat{\Xi} f(r) = \sigma f(r) \ \Longrightarrow \ f(r) \propto r^\sigma $$
    for a constant exponent $\sigma$. For a power-law $f(r)$, Eq. \eqref{eq:from_f_to_F} becomes
    $$F(r,b) = M(b) f(g(b)r) = M(b) g^\sigma(b) f(r) \ , $$
    which means the \textit{strength renormalization} and \textit{radial rescaling} are non-separable, therefore this case has already been considered within the \textit{spherical property}, i.e. by fixing $g(b)=1$ and allowing $M(b)$ to vary.
\end{itemize}

Another case not mentioned in the main text could arise in Eq. \eqref{eq:2power} when $\sigma_+=\sigma_- \equiv \sigma$, in which the general solution then takes the form:
$$ f(r) = r^\sigma \Big( U_1 + U_2 \ln(r) \Big) \ . $$
However, this force field only satisfies Eq. \eqref{eq:perturb} when $U_2 = 0$, reducing it to a power-law which has already been considered.

\section{Multipolar Potential \label{app:covariant_multipole}}

Let us precisely define a point-source multipole in flat space $\mathbb{R}^n$ using a concrete distribution density model. A unit point monopole located at the origin is represented by the
$n$-dimensional delta-function density:
\begin{equation}
    \rho_0(\vec{r})
    =
    \delta^{(n)}(\vec{r})\ ,
\end{equation}
in which the delta-function is defined via the integration property \cite{hilbert1985methods}:
$$
    \int_{\mathbb{R}^n}
    d^n\vec{r} \,
    \delta^{(n)}(\vec{r})\,
    h(\vec{r})
    =
    h(0) $$
for any smooth scalar function $h(\vec{r})$, meaning that the entire source strength is concentrated at the origin $\vec{r}=0$. A \textit{higher-rank multipole} can be obtained by considering a localized
arrangement of monopolar point-sources (i.e. the \textit{indistinguishability} between Fig. \ref{fig01}a2 and b2) whose \textit{lower-order moments cancel}, while a
\textit{higher-order moment remains finite} as the size of the arrangement is
taken to zero. In this point-like limit, the resulting source has no net
monopole strength and, for a pure rank-$\ell$ multipole, no moments of
rank lower than $\ell$. Derivatives of the delta function provide a
natural distributional representation of precisely this limit, i.e.
$$
\int_{\mathbb R^n}
d^n\vec r \ 
\partial_{i_1} \partial_{i_2} ... \partial_{i_\ell}
\delta^{(n)}(\vec r) 
h(\vec r)
=
(-1)^\ell 
\partial_{i_1} \partial_{i_2} ... \partial_{i_\ell}
h(0) \ , 
$$
where $\partial_i \equiv \partial/\partial r_i$ and $i_1, i_2, ..., i_\ell$ are the Cartesian indices ranging from $1$ to $n$ \cite{kocher1978point}. Since spatial derivatives commute, this tensor is automatically symmetric in its indices. To obtain a pure rank-$\ell$ multipole, we further remove the trace components by taking the symmetric-traceless part. The resulting distribution density represents a component of an irreducible rank-$\ell$ multipole:
$$ \partial_{\langle i_1} \partial_{i_2}  ... \partial_{i_\ell \rangle} \delta^{(n)}(\vec{r}) \ , $$
where $\langle ... \rangle$ denotes symmetrization over the enclosed indices
followed by the removal of all traces. Adding up all components, we arrive at a general distribution density model for a pure rank-$\ell$ point-source
multipole:
$$\rho_{\ell}(\vec{r})
    = \sum_{i_1,i_2,...,i_\ell} Q_{i_1 i_2 ... i_\ell}
    \partial_{\langle i_1} \partial_{i_2}  ... \partial_{i_\ell \rangle} 
    \delta^{(n)}(\vec{r}) \ , $$
where the tensor
$Q_{i_1 i_2 ... i_\ell}$ specifies the multipole
strength and orientation. The factor
$(-1)^\ell/\ell!$ is conventional and may instead be absorbed into
the definition of $Q_{i_1 i_2 ... i_\ell}$. The total potential created by this at a reference point $\vec{a}$ is given by:
$$
    \phi_\ell (\vec{a})
    =
    \int
    d^n \vec{r} \ 
    \rho_\ell (\vec{r})
    \phi\big(|\vec{a}-\vec{r}|\big) =
    \sum_{i_1,i_2,...,i_\ell} Q_{i_1 i_2 ... i_\ell}
    \partial_{\langle i_1} \partial_{i_2}  ... \partial_{i_\ell \rangle} \phi(a) \ \ \text{where} \ \ a \equiv |\vec{a}| \ ;
$$
note that the derivative operator $\partial_i$ in the definition of $\rho_\ell(\vec r)$ acts with respect to the source coordinates $r_i$, whereas after evaluating the distributional integral, the corresponding derivative on the potential acts with respect to the reference coordinates $a_i$, hence there is no $(-1)^\ell$-factor. Applying Hobson's differentiation formula \cite{nomura2018proof}:
\begin{equation}
    \partial_{\langle i_1} \partial_{i_2} ...
    \partial_{i_\ell\rangle}\phi(a)
    =
    a^\ell
    \left(
        \frac{1}{a}\frac{d}{da}
    \right)^\ell
    \phi(r)\,
    \vec{e}_{\langle i_1} \vec{e}_{i_2} ... \vec{e}_{i_\ell\rangle} \ ,
\end{equation}
where $\vec{e}_i=a_i/a$, we obtain
$$\phi_{\ell}(\vec{a})
    =
    \psi_\ell(a)\mathcal Y_\ell(\Omega) \ , $$
with
$$\psi_\ell(a)
    =
    \hat{D}_\ell\phi(a) \ \ \text{where} \ \
    \hat{D}_\ell
    \equiv
    a^\ell
    \left(
        \frac{1}{a}\frac{d}{da}
    \right)^\ell \ ,
$$
which is identical to Eq. \eqref{eq:multipole_operator}.
The orientation feature
$$\mathcal Y_\ell(\Omega)
    \equiv 
    \sum_{i_1,i_2,...,i_\ell} Q_{i_1 i_2 ... i_\ell}
    \vec{e}_{\langle i_1} \vec{e}_{i_2} ... \vec{e}_{i_\ell\rangle} 
$$
can be rewritten as sum of hyperspherical-harmonic modes
$$ \mathcal Y_\ell(\Omega)
    = \sum_{\mathbf{m}} c_{\ell \mathbf{m}} Y_{\ell \mathbf{m}}(\Omega) \ , $$
where the coefficients $c_{\ell \mathbf{m}}$ can be identified with:
$$  c_{\ell \mathbf{m}} \equiv Q_{i_1 i_2 ... i_\ell} \sum_{i_1,i_2,...,i_\ell} \oint d\Omega  \ Y_{\ell \mathbf{m}}^*(\Omega)
    \vec{e}_{\langle i_1} \vec{e}_{i_2} ... \vec{e}_{i_\ell\rangle}  \ . $$

\section{Angular Structure under Spherical Mean \label{app:angular_kept}}

For simplicity, we first consider $n=3$:
$$ \text{Vol}\left[ S^{n-1}(b)\right] = 4\pi b^2 \ \ \text{and} \ \ d\sigma(b\vec{e}_s) = b^2 \sin\theta_s d\theta_s d\varphi_s \ , $$
where $(\theta_s,\varphi_s)$ encodes the spherical angles on the spherical shell. Consider a single $m$-component of the rank-$\ell$ with unit strength, i.e.
\begin{equation}
    \phi_\ell(\vec{r}) \rightarrow \phi_{\ell m} = \psi_\ell (r) Y_{\ell m}(\Omega) \ ,
    \label{eq:oh}
\end{equation}
then the spherical mean in Eq. \eqref{eq:spherical_mean_ell} can be calculated from:
$$ \Phi_{\ell m}(\vec{r},b) = \frac1{4\pi} \int^{2\pi}_0 d\varphi_s \int^\pi_0 d\theta_s \ \sin\theta_s \psi_\ell(r_*) Y_{\ell m}(\Omega_*) \ , $$
where
$\vec{r}_*=\vec{r}-b\vec{e}_s \equiv (r_*,\Omega_*)$. 

Consider the potential at another point $\vec{r}_{R}$ which is related to $\vec{r}$ by a rotation $R(...)$, i.e. look at point $R(\vec{r})$, then:
\begin{equation}
    \Phi_{\ell m}\left(R(\vec{r}),b\right) = \frac1{4\pi} \int^{2\pi}_0 d\varphi_s \int^\pi_0 d\theta_s \ \sin\theta_s \psi_\ell(|R(\vec{r}) - b\vec{e}_s|) Y_{\ell m}(\Omega_{new}) \ ,
    \label{eq:int_rot}
\end{equation}
where $\Omega_{new}$ is the direction of $R(\vec{r})-b\vec{e}_s$. We now change the integration variable from $\vec e_s$ to $\vec e_{**}$, defined by
$$\vec e_s=R\left(\vec e_{**}\right) \ , $$
and let us denote $(\theta_{**},\varphi_{**})$ to be the spherical angles of the direction $\vec{e}_{**}$ on the spherical shell. We also define $\vec{r}_{**} = \vec{r} - b \vec{e}_{**} \equiv (r_{**},\Omega_{**})$, then
$$R(\vec{r}) - b\vec{e}_{s} = R(\vec{r}) - b R(\vec{e}_{**}) = R(\vec{r}-b\vec{e}_{**})= R(\vec{r}_{**}) \ , $$
therefore $|R(\vec{r}) - b\vec{e}_{s}|= |R(\vec{r}_{**})|\equiv r_{**}$ and $\Omega_{new}$ is the direction of $R(\vec{r}_{**})$, i.e. $\Omega_{new} = R\Omega_{**}$. The spherical harmonics transforms under a rotation as \cite{varshalovich1988quantum}:
$$ Y_{\ell m}(\Omega_{new}) = Y_{\ell m}(R\Omega_{**}) = \sum^{\ell}_{m_{**} = -\ell} D^{(\ell)}_{m_{**} m}(R) Y_{\ell m_{**}} (\Omega_{**})  \ , $$
where $D^{(\ell)}$ is the Wigner $D$-matrix and $D^{(\ell)}_{m_{**} m}(R)$ is a number determined by the rotation $R(...)$. We can then rewrite the integration in Eq. \eqref{eq:int_rot} as:
\begin{equation}
\begin{split}    \Phi_{\ell m}\left(R(\vec{r}),b\right) &= \frac1{4\pi} \int^{2\pi}_0 d\varphi_{**} \int^\pi_0 d\theta_{**} \ \sin\theta_{**} \psi_\ell(r_{**}) \sum^{\ell}_{m_{**} = -\ell} D^{(\ell)}_{m_{**} m}(R) Y_{\ell m_{**}} (\Omega_{**}) 
\\
&= \sum^{\ell}_{m_{**} = -\ell} D^{(\ell)}_{m_{**} m}(R) \left[ \frac1{4\pi} \int^{2\pi}_0 d\varphi_{**} \int^\pi_0 d\theta_{**} \ \sin\theta_{**} \psi_\ell(r_{**}) Y_{\ell m_{**}} (\Omega_{**}) \right]
\\
& = \sum^{\ell}_{m_{**} =-\ell} D^{(\ell)}_{m_{**} m}(R) \Phi_{\ell m_{**}} (\vec{r},b) \ ,
\end{split}
\label{eq:rot_transf}
\end{equation}
where to go from the second-line to the third-line, note that the integration in $[...]$ is exactly the same as Eq. \eqref{eq:oh} but with $m \rightarrow m_{**}$ and a change of variable $\vec{e}_s \rightarrow \vec{e}_{**}$. 

Eq. \eqref{eq:rot_transf} implies that $\Phi_{\ell m}(\vec{r},b)$ transforms under rotations in the same way as the spherical harmonics $Y_{\ell m}(\Omega)$. By Schur's lemma \cite{hall2013lie}, it therefore separates into a radial factor and the angular part $Y_{\ell m}(\Omega)$, i.e.
$$\Phi_{\ell m}(\vec{r},b) \propto Y_{\ell m}(\Omega) \ , $$
and, upon summing over all components from the same rank-$\ell$, we arrive at Eq. \eqref{eq:get_F}.

For a general dimension $n$, similar arguments follow. For each fixed $r$, the spherical-mean operator commutes with the action of the rotation group SO$(n)$. The degree-$\ell$ hyperspherical harmonics form an irreducible representation of SO$(n)$. Therefore, by Schur's lemma, the mean operator acts as a scalar on this angular subspace and preserves the angular factor $\mathcal Y_\ell(\Omega)$.

\section{Derivation of Eq. \eqref{eq:multipolar_rel} \label{app:multipolar_cond}}

From the relation between force field and interaction potential of a multipolar point-source, we can apply the chain rule to obtain:
\begin{equation}
\begin{split}
    \vec f_{\ell} \left(g_{\ell}(b)\vec r\right)
    &= -\vec{\nabla}_{\vec{a}} \phi_\ell(\vec{a})\Big|_{\vec{a}=g_\ell(b)\vec{r}} = -\frac{1}{g_{\ell}(b)}\vec\nabla_{\vec r} \phi_\ell \left( g_\ell (b) \vec{r} \right) 
    \\
    &= -\frac{1}{g_{\ell}(b)}\vec\nabla_{\vec r}
      \Big[
        \psi_{\ell} \left(g_{\ell}(b)r\right) \mathcal Y_{\ell}(\Omega)
      \Big] \ ,
\end{split}
\end{equation}
where we have used Eq. \eqref{eq:curved_point_multipole} to arrive at the second-line. We use this and Eq. \eqref{eq:multipole_force_pot} to rewrite Eq. \eqref{eq:multipolar-shape-preserving-condition} as: 
\begin{equation*}
    \vec{\nabla}_{\vec{r}} \Phi_{\ell}(\vec{r},b)
    =-M_{\ell}(b)\,
      \vec f_{\ell} \left(g_{\ell}(b)\vec r\right) =  -\frac{M_\ell(b)}{g_{\ell}(b)}\vec\nabla_{\vec r}
      \Big[
        \psi_{\ell} \left(g_{\ell}(b)r\right) \mathcal Y_{\ell}(\Omega)
      \Big] \ ,
\end{equation*}
which can be reorganized into:
\begin{equation*}
    \vec\nabla_{\vec r}
    \left\{
      \left[
        \Psi_{\ell}(r,b)
        -\frac{M_{\ell}(b)}{g_{\ell}(b)}
         \psi_{\ell} \left(g_{\ell}(b)r\right)
      \right] \mathcal Y_{\ell}(\Omega)
    \right\}=0.
\end{equation*}
Since $\mathcal Y_{\ell}(\Omega)$ is non-constant for $\ell>0$, the expression in
$[...]$ must vanish.  Thus we arrive at:
\begin{equation}
    \Psi_{\ell}(r,b)
    = N_\ell (b)\psi_{\ell} \left(g_{\ell}(b)r\right) \ \ \text{where} \ \ N_\ell(b) \equiv    \frac{M_{\ell}(b)}{g_{\ell}(b)} \ ,
\end{equation}
which is identical to Eq. \eqref{eq:multipolar_rel}. The boundary conditions, in which the spherical shell becomes a point-source, at $b=0$ must be:
\begin{equation*}
M_\ell (0)=1 \ , \ g_\ell (0)=1 \ ,  
\end{equation*}
which also implies $N_\ell (0)=1$. The integration Eq. \eqref{eq:spherical_mean_ell} for $\Phi_\ell(\vec{r},b)$ is invariant under the transformation $b \rightarrow -b$, therefore we can also set:
\begin{equation*}
    M_\ell'(0)=0 \ , \ g_\ell'(0)=0 \ ,
\end{equation*}
which also implies $N_\ell'(0)=0$.

\section{Derivation of Eq. \eqref{eq:multipolar_local_constraint} and Eq. \eqref{eq:multipolar_master_equation}  \label{app:derive_multipolar}}

We follow similar steps as in Appendix \ref{app:ff_ode}. Define:
$$ a \equiv g_{\ell}(b)r \ , \ \mu_{\ell}(b)\equiv\frac{N_{\ell}'(b)}{N_{\ell}(b)} \ , \ \gamma_{\ell}(b)\equiv\frac{g_{\ell}'(b)}{g_{\ell}(b)} \ . $$
Substituting Eq. \eqref{eq:multipolar_rel} into the EPD
identity Eq. \eqref{eq:EPD_rank_ell} yields \eqref{eq:multipolar_master_equation}, i.e.
\begin{equation}
    V_1(b)\psi_{\ell}(a)
    +V_2(b)\hat\Xi\psi_{\ell}(a)
    +{g'_{\ell}}^{2}(b)\hat\Xi^{2}\psi_{\ell}(a)
    =g_{\ell}^{4}(b)\hat L_{\ell}\psi_{\ell}(a) \ ,
    \label{eq:riposa_in_pace}
\end{equation}
where
\begin{equation*}
\begin{split}
    V_1(b)
    &\equiv
      g^2_\ell(b) \left[ \mu_{\ell}'(b)+\mu_{\ell}^{2}(b)
      +\frac{n-1}{b}\mu_{\ell}(b) \right]\ ,
      \\
    V_2(b)
    &\equiv
      g_\ell^2(b) \left[ \gamma_{\ell}'(b)
      +2\mu_{\ell}(b)\gamma_{\ell}(b)
      +\frac{n-1}{b}\gamma_{\ell}(b) \right] \ .
\end{split}
\end{equation*}

If Eq. \eqref{eq:multipolar_rel} holds for all values of $b$, it must also hold for infinitesimal values of b in a
neighborhood of $b=0$.  Consider the following Taylor expansions:
\begin{equation*}
    \begin{split}
        \Psi_{\ell}(r,b)
    &\approx \psi_{\ell}(r)
      +\frac{b^{2}}{2n}\hat L_{\ell}\psi_{\ell}(r) \ ,
      \\
    N_{\ell}(b)\psi_{\ell}\left(g_{\ell}(b)r\right) 
    & \approx \psi_{\ell}(r)
      +\frac{b^{2}}{2}
       \left[
         N_{\ell}''(0)\psi_{\ell}(r)
         +g_{\ell}''(0)\hat\Xi\psi_{\ell}(r)
       \right] \ .
    \end{split}
\end{equation*}
Matching the terms of order $b^{2}$ gives
\begin{equation*}
    \hat L_{\ell}\psi_{\ell}(r)
    =\alpha_{\ell}\psi_{\ell}(r)
     +\beta_{\ell}\hat\Xi\psi_{\ell}(r),
\end{equation*}
where $\alpha_{\ell}$ and $\beta_{\ell}$ are two constants
\begin{equation*}
    \alpha_{\ell}
    \equiv n N_{\ell}''(0)
    =n\left[M_{\ell}''(0)-g_{\ell}''(0)\right] \ , \ 
    \beta_{\ell}\equiv ng_{\ell}''(0) \ ,
\end{equation*}
which is the same as Eq. \eqref{eq:multipolar_local_constraint}. We can then use this expression of $\hat{L}_\ell \psi_\ell(r)$ to rewrite Eq. \eqref{eq:riposa_in_pace} as Eq. \eqref{eq:multipolar_master_equation}, in which
$$W_1 \equiv \alpha_\ell g^4_\ell(b) - V_1(b) \ , \ W_2 \equiv \beta_\ell g_\ell^4(b) - V_2(b) \ . $$

\section{Multipolar Spherical Property \label{app:multipolar_sp}}

For the \textit{multipolar spherical property} to emerge, the exterior force field generated by a point multipole at the center must be identical, up to a shell-\textit{radius-dependent rescaling factor} $M_\ell(b)$, to that generated by uniformly smearing the same aligned multipole over $S^{n-1}(b)$:
\begin{equation}
    -\vec{\nabla}_{\vec{r}} \Phi_\ell (\vec{r},b) =  - M_\ell(b) \vec{\nabla}_{\vec{r}} \varphi_\ell (\vec{r}) \ .
    \label{eq:force_field}
\end{equation} 
We exclude $\ell=0$ from further analysis, as this case reduces precisely to the study of \textit{monopolar spherical property} established in our previous work \cite{tse2026generalizing} -- we summarize all relevant findings there in Appendix \ref{app:monopolar}. 

For $\ell > 0$, Eq. \eqref{eq:curved_point_multipole} and Eq. \eqref{eq:get_F} allow us to rewrite Eq. \eqref{eq:force_field} as:
$$\vec{\nabla}_{\vec{r}} \left\{ \Big[ \Psi_\ell(r,b) - M_\ell(b) \psi_\ell(r) \Big] \mathcal{Y}_\ell(\Omega) \right\} = 0 \ . $$
Since $\ell>0$, the angular function $\mathcal{Y}_\ell(\Omega)$ is non-constant, hence the angular component of the left-hand side (LHS) therefore vanishes everywhere only if
\begin{equation}
        \Psi_\ell(r,b)=M_\ell(b) \psi_\ell(r) \ .
        \label{eq:cond_on_F}
\end{equation}
Substituting this into the preceding equation makes it vanish identically, showing that this condition is both \textit{necessary} and \textit{sufficient}.

Replace the expression for $\Psi_\ell(r,b)$ from Eq. \eqref{eq:cond_on_F} into the PDE constraint Eq. \eqref{eq:EPD_rank_ell}:
\begin{equation}
    -\lambda(b) =  \frac{\psi_\ell''(r) + \frac{(n-1)}{r}  \psi'_\ell(r) - \frac{\Lambda_\ell}{r^2} \psi_\ell(r)}{\psi_\ell(r) } \ ,
    \label{eq:solve_for_f}
\end{equation}
where $\Lambda_\ell = \ell(\ell + n-2)$ and 
\begin{equation}
    \lambda(b) \equiv -\frac{ M''_\ell(b) + \frac{(n-1)}{b} M'_\ell(b)}{M_\ell(b)} \ .
    \label{eq:lambda_from_M}
\end{equation}
Since the right-hand side (RHS) of Eq. \eqref{eq:solve_for_f} does not depend on $b$, the LHS should also not, therefore $\lambda(b)$ must be a constant, i.e. $\lambda(b)=\lambda$. For the special case $M_\ell(b)=1$, the force fields are identical, i.e. Eq. \eqref{eq:force_field} becomes
$$ -\vec{\nabla}_{\vec{r}} \Phi_\ell (\vec{r},b) =  - \vec{\nabla}_{\vec{r}} \phi_\ell (\vec{r}) \ , $$
and the \textit{multipolar spherical property} reduces to the \textit{multipolar shell theorem}. From Eq. \eqref{eq:lambda_from_M}, this case corresponds to
\begin{equation}
    \lambda=0 \ .
    \label{eq:shell_theorem}
\end{equation}

For a constant value of $\lambda$, we can solve the ODE in Eq. \eqref{eq:solve_for_f} to obtain the radial function $\psi_\ell(r)$ that gives rise to the \textit{spherical property}, which can be divided into two different cases:
\begin{itemize}
    \item \underline{$\lambda \neq 0$}: For $\lambda > 0$, the general solution is given by a linear combination of the Bessel functions: 
    $$
        \psi_\ell (r) = r^{-\alpha} \left[ A_1 J_{\ell + \alpha}(\kappa r) + A_2  Y_{\ell+\alpha}(\kappa r) \right] \ ,
    $$
    where $\alpha = n/2 - 1$, $\kappa=\lambda^{1/2}$, and the $A$-values are arbitrary constants. The corresponding potential, as follows from Eq. \eqref{eq:multipole_operator}, has the following general form:
    \begin{equation}
        \phi (r) = r^{-\alpha} \left[ B_1 J_{\alpha}(\kappa r) + B_2 Y_{\alpha}(\kappa r) \right] + \sum^{\ell-1}_{j=0} C_j r^{2j}  \ ,
        \label{eq:lambda_not_0}
    \end{equation}
    where we absorb any overall factor outside of the $A$-coefficients into the $B$-coefficients and the $C$-values are arbitrary constants. The \textit{strength renormalization} factor $M_\ell(b)$ can be found by solving Eq. \eqref{eq:lambda_from_M} subject to the boundary conditions $M_\ell(0)=1$ and $M_\ell'(0)=0$, which gives:
    $$M_\ell(b) = \Gamma\left(\frac{n}2 \right) \left( \frac{2}{\kappa b} \right)^\alpha J_\alpha (\kappa b) \ , $$
    which is independent of the rank-$\ell$ and also identical to the factor from the \textit{monopolar spherical property} Eq. \eqref{eq:mono_M}. For $\lambda < 0$, the answer can be obtained by analytic continuation

\item \underline{$\lambda = 0$}: This is the situation of Eq. \eqref{eq:shell_theorem}, where we obtain the \textit{multipolar shell theorem}. The general solution of Eq. \eqref{eq:solve_for_f}  for $\psi_\ell(r)$ is given by:
\begin{equation}
    \psi_\ell(r) = \frac{A_3}{r^{\ell + n - 2}} + A_4 r^\ell \ ,
\end{equation}
where the $A$-values are arbitrary constants. The corresponding potential, as follows from Eq. \eqref{eq:multipole_operator}, has the following general form:
\begin{equation}
\phi(r)=
\begin{cases}
B_3 \ln (r)  + B_4 r^{2\ell}+\displaystyle\sum_{j=0}^{\ell-1}D_j r^{2j}
& \text{for} \ \ n=2 \ , \\[16pt]
\displaystyle{\frac{B_3}{r^{n-2}}} +B_4 r^{2\ell}
+\displaystyle\sum_{j=0}^{\ell-1}D_j r^{2j}
& \text{for} \ \ n>2 \ ,
\end{cases}
\label{eq:lambda_is_0}
\end{equation}
where we absorb any overall factor outside of the $A$-coefficients into the $B$-coefficients and the $D$-values are arbitrary constants.
\end{itemize}

It is important to note that these interactions possessing \textit{multipolar spherical property} admit an additional polynomial potential between rank-$(\ell-1)$ and rank-$\ell$, i.e. a contribution $
\propto r^{2(\ell-1)}$, which, despite being nontrivial, generates no rank-$\ell$ multipole field as follows from Eq. \eqref{eq:multipole_operator}. In other words, these potentials are \textit{invisible} to the rank-$\ell$ multipole field, as explained mathematically by Eq. \eqref{eq:kernel}. Beyond these potentials, compared to the \textit{monopolar spherical property} (see Appendix \ref{app:monopolar}), the only new potentials at rank-$\ell$ are those proportional to $\propto r^{2\ell}$, i.e.
$$\hat{D}_\ell r^{2\ell} = 2^\ell \ell ! r^\ell \neq 0 \ . $$

\end{document}